\documentclass[12pt]{ussci}
\usepackage[T1]{fontenc}
\usepackage{array}
\usepackage{amsmath}
\usepackage{amsfonts}
\usepackage{amssymb}
\usepackage{amsbsy}%
\usepackage{graphicx}
\usepackage{setspace}
\usepackage{multirow}
\usepackage{bm}
\usepackage[mathscr]{euscript}
\usepackage[backend=biber]{biblatex}
\newcommand\papertopic{Other}

\title{ Direct numerical simulation of high-pressure mixing in turbulent jets }

\author[1]{Nek Sharan}
\author[1,2,*]{Josette Bellan}

\affil[1]{Mechanical and Civil Engineering, California Institute of Technology, Pasadena, CA 91125, USA}
\affil[2]{Jet Propulsion Laboratory, California Institute of Technology, Pasadena, CA 91109, USA}
\affil[*]{Corresponding author: \email{josette.bellan@jpl.nasa.gov}}

\begin{document}
\maketitle

\begin{abstract} 
Combustion in automotive and aerospace applications employing diesel,
gas turbine and liquid rocket engines is preceded by injection and
mixing of fuel and oxidizer at high pressures, often exceeding mixture
critical values. Experimental observations indicate that the jets
injected at supercritical pressures exhibit significantly different
dynamics than the jets at subcritical conditions, owing to the lack
of distinct liquid and gas phases in supercritical state. As a result,
the averaged flow quantities such as the potential core length, jet
spatial growth rate and velocity decay profiles differ in the two
conditions, resulting in different mixed-fluid distributions. In this
study, turbulent jet direct numerical simulations (DNS) are performed
to examine the variations in statistics between injection of Nitrogen
($\mathrm{N_{2}}$) in Nitrogen ($\mathrm{N_{2}}$) at subcritical
(perfect-gas) and supercritical conditions. Isothermal round jets
at Reynolds number ($Re_{D}$), based on jet diameter ($D$) and jet
orifice velocity ($U_{0}$), of $5000$ and Mach number of $0.6$
are considered. For mixing analyses, a passive scalar transported
with the flow is examined.
\end{abstract}



\section{Introduction\label{sub:Introduction}}

Fuel injection and turbulent mixing at supercritical pressures determines
ignition and combustion in numerous engineering applications. Flow
evolution under such conditions is characterized by strong non-linear
coupling between dynamics, transport coefficients, and thermodynamics.
A model that accounts for these non-linear effects in computation
of the thermodynamic state of the mixture, and the heat and mass fluxes
in a multi-component fluid-flow simulation was proposed by Masi \textit{et
al.} \cite{masi2013multi}. The goal of the present study is to evaluate
the model in a turbulent free-jet configuration to simulate fuel injection
and mixing in high-pressure($p$) combustion chambers of propulsion
systems. 

Validation of numerical simulations at high-$p$ conditions, where
theoretical results are scarce, requires comparisons with experimental
data. However, most experimental studies (\textit{e.g.}, \cite{chehroudi2002visual,roy2013disintegrating,muthukumaran2016mixing})
inject liquid (fuel at subcritical conditions) or high-density fluid
(fuel at supercritical conditions) at high Reynolds numbers ($Re_{D}\geq20,000$),
and provide qualitative visual information about flow dynamics and
mixing. A numerical simulation of these flows, for direct comparisons
with the experiments, would require several models. For example, models
to accommodate potential two-phase/density-jump regions and to account
for the subgrid-scale fluxes would be required, in addition to the
thermodynamic model. Moreover, appropriate inflow and boundary treatments
that accurately replicate the experimental conditions are also imperative.
Interactions between the models and numerical details make validation
of individual models infeasible in such complex flows at high-$Re_{D}$.
Moreover, the lack of quantitative turbulence/mixing statistics measurements
in high-$p$ experiments makes assessment of a model even more challenging. 

To mitigate the model interactions, direct numerical simulations at
$Re_{D}$ of $5,000$ are considered in this study. Jet flow at perfect-gas
conditions, for which theoretical \cite{morris1983viscous,michalke1984survey}
and experimental \cite{panchapakesan1993turbulence,hussein1994velocity}
results exist, is first considered to validate the numerical setup
and to create a database for flow-behavior comparisons against multicomponent
flows at high pressures. To systematically initiate the study, mixing
behavior in the single-species flows is assessed by a virtual passive
scalar transported with the flow, modeling diffusion at unity Schmidt
number ($Sc$) justifiable under perfect-gas conditions.

Accurate turbulent free-jet flow computation requires a careful choice
of inflow/boundary conditions, domain size, and numerical discretization.
The near-field jet flow evolution is particularly sensitive to the
choice of inflow perturbations, and several studies \cite{boersma1998numerical,bogey2005effects}
have examined its influence on turbulence statistics. The jet flow
attains self-similarity after the development region, containing potential
core collapse and transition to turbulence; the axial distance to
attain a self-similar state depends on the inflow perturbations. Moreover,
theoretical results \cite{george1989self} suggest that the self-similar
state may also depend on the inflow, requiring the similarity variables
to be appropriately modified for analyses. Although the theoretical
and experimental jet flow studies focus on the flow statistics in
the self-similar region, in practical applications with small combustion
chambers, the jet near-field is equally important because this is
where the phenomena determining the flame occur, and this is where control
must be exercised to influence flow behavior.

\section{Governing equations\label{sub:governing_eqn}}

For the single-species flow of interest here, the conservation equations solved in this study are:
\begin{equation}
\frac{\partial\rho}{\partial t}+\frac{\partial}{\partial x_{j}}\left[  \rho
u_{j}\right]  =0, \label{eq:eq1}%
\end{equation}%
\begin{equation}
\frac{\partial}{\partial t}\left(  \rho u_{i}\right)  +\frac{\partial
}{\partial x_{j}}\left[  \rho u_{i}u_{j}+p\delta_{ij}-\sigma_{ij}\right]  =0,
\label{eq:eq2}%
\end{equation}%
\begin{equation}
\frac{\partial}{\partial t}\left(  \rho e_{t}\right)  +\frac{\partial
}{\partial x_{j}}\left[  \left(  \rho e_{t}+p\right)  u_{j}-u_{i}\sigma
_{ij}+q_{j}\right]  =0, \label{eq:eq3}%
\end{equation}%
\begin{equation}
\frac{\partial}{\partial t}\left(  \rho \xi\right)  +\frac{\partial
}{\partial x_{j}}\left[  \rho \xi\ u_{j}+J_{j}\right]  =0,
\label{eq:eq4}%
\end{equation}
where $t$ denotes the time, $x$ is a Cartesian
coordinate, subscripts $i$ and $j$ refer to the spatial coordinates, $u_{i}$
is the velocity, $p$ is the pressure, $\delta_{ij}$ is the Kronecker delta, 
$e_{t}=e+u_{i}u_{i}/2$ is the total energy (\textit{i.e.}, internal
energy, $e$, plus kinetic energy), $\xi\in\left[0,1\right]$ is a virtual passive scalar
transported with the flow, $\sigma_{ij}$ is the Newtonian viscous stress tensor%
\begin{equation}
\sigma_{ij}=\mu\left(  2S_{ij}-\dfrac{2}{3}S_{kk}\delta_{ij}\right)  ,\text{
\ \ }S_{ij}=\dfrac{1}{2}\left(  \dfrac{\partial u_{i}}{\partial x_{j}}%
+\dfrac{\partial u_{j}}{\partial x_{i}}\right)  , \label{eq:stress}%
\end{equation}
where $\mu$ is the viscosity, $S_{ij}$ is the strain-rate tensor, and
$q_{j}=-\lambda\thinspace\partial T/\partial x_{j}$ and $J_{j}=-\mathscr{D}\thinspace\partial\xi/\partial x_{j}$ 
are the $j$-direction heat flux and scalar diffusion flux, respectively. $\lambda$ 
is the thermal conductivity and $\mathscr{D}=\mu/Sc$ is the scalar diffusivity,
where $Sc$ denotes the Schmidt number. The injected fluid is assigned a 
scalar value of 1, whereas the chamber fluid a value of 0.

Two jet-flow simulations at conditions summarized in Table \ref{tab:Summary_of_cases}
are performed to examine flow statistics differences between injection
at perfect-gas and supercritical conditions. Only the chamber pressure
$p_{\infty}$ differs between the two cases.

For the near-atmospheric-$p$ simulation (Case 1), the perfect gas
equation of state is applicable, given by
\[
p=\frac{\rho R_{\mathrm{u}}T}{m},
\]
where $R_{u}$ is the universal gas constant and $m$ is the species
molar mass. The viscosity is modeled as a power law
\[
\mu=\mu_{R}\left(\frac{T}{T_{R}}\right)^{n}
\]
with $n=2/3$ and the reference viscosity $\mu_{R}=\rho_{0}U_{0}D/Re_{D}$,
where $\rho_{0}$ and $U_{0}$ are the jet-exit fluid density and
velocity, respectively, and the reference temperature $T_{R}=293\thinspace\mathrm{K}$.
The thermal conductivity $\lambda=\mu C_{p}/\mathrm{Pr}$, where Prandtl
number $\mathrm{Pr}=0.7$, the ratio of specific heats $\gamma=1.4$,
and the isobaric heat capacity $C_{p}=\gamma R_{\mathrm{u}}/\left(\gamma-1\right)$
is assumed.

For the high-$p$ simulation (Case 2), the governing equations (\ref{eq:eq1})-(\ref{eq:eq4})
are closed using the Peng-Robinson (PR) equation of state (EOS)
\[
p=\frac{R_{\mathrm{u}}T}{\left(v_{\mathrm{PR}}-b_{\mathrm{mix}}\right)}-\frac{a_{\mathrm{mix}}}{\left(v_{\mathrm{PR}}^{2}+2b_{\mathrm{mix}}v_{\mathrm{PR}}-b_{\mathrm{mix}}^{2}\right)},
\]
where the pressure, $p$, and temperature, $T$, are obtained as an
iterative solution from the density, $\rho$, and internal energy,
$e$, obtained from the conservation equations \cite{okong2002directAIAA}.
The molar PR volume $v_{\mathrm{PR}}=v-v_{\mathrm{s}}$, where the
molar volume $v=m/\rho$. $v_{\mathrm{s}}$ denotes the volume shift
introduced to improve the accuracy of the PR EOS at high pressures
\cite{okong2002directAIAA,harstad1997efficient}. $a_{\mathrm{mix}}$
and $b_{\mathrm{mix}}$ are obtained from the expressions detailed
in \cite[Appendix B]{sciacovelli2019influence}.

The physical viscosity, $\mu_{\mathrm{ph}}$, and thermal conductivity,
$\lambda_{\mathrm{ph}}$, are calculated using the Lucas method \cite[Chapter 9]{poling2001properties}
and the Stiel-Thodos method \cite[Chapter 10]{poling2001properties},
respectively. The computational viscosity, $\mu$, and thermal conductivity,
$\lambda$, are then obtained by scaling $\mu_{\mathrm{ph}}$ and
$\lambda_{\mathrm{ph}}$ with factor $\mathcal{F}=\mu_{R}/\mu_{\mathrm{ph},0}$,
\textit{i.e.}, $\mu=\mathcal{F}\mu_{\mathrm{ph}}$ and $\lambda=\mathcal{F}\lambda_{\mathrm{ph}}$,
to simulate the flow at a specified Reynolds number $Re_{D}$. The
inflow physical viscosity, $\mu_{\mathrm{ph},0}$, is obtained from
the Lucas method using the pressure $p_{\infty}$ and the average
temperature $\left(T_{\mathrm{inj}}+T_{\mathrm{ch}}\right)/2$, where
the subscripts ``$\mathrm{inj}$'' and ``$\mathrm{ch}$'' denote
the injection and chamber conditions, respectively.

\begin{figure}[!htb]
\begin{centering}
\includegraphics[width=9cm]{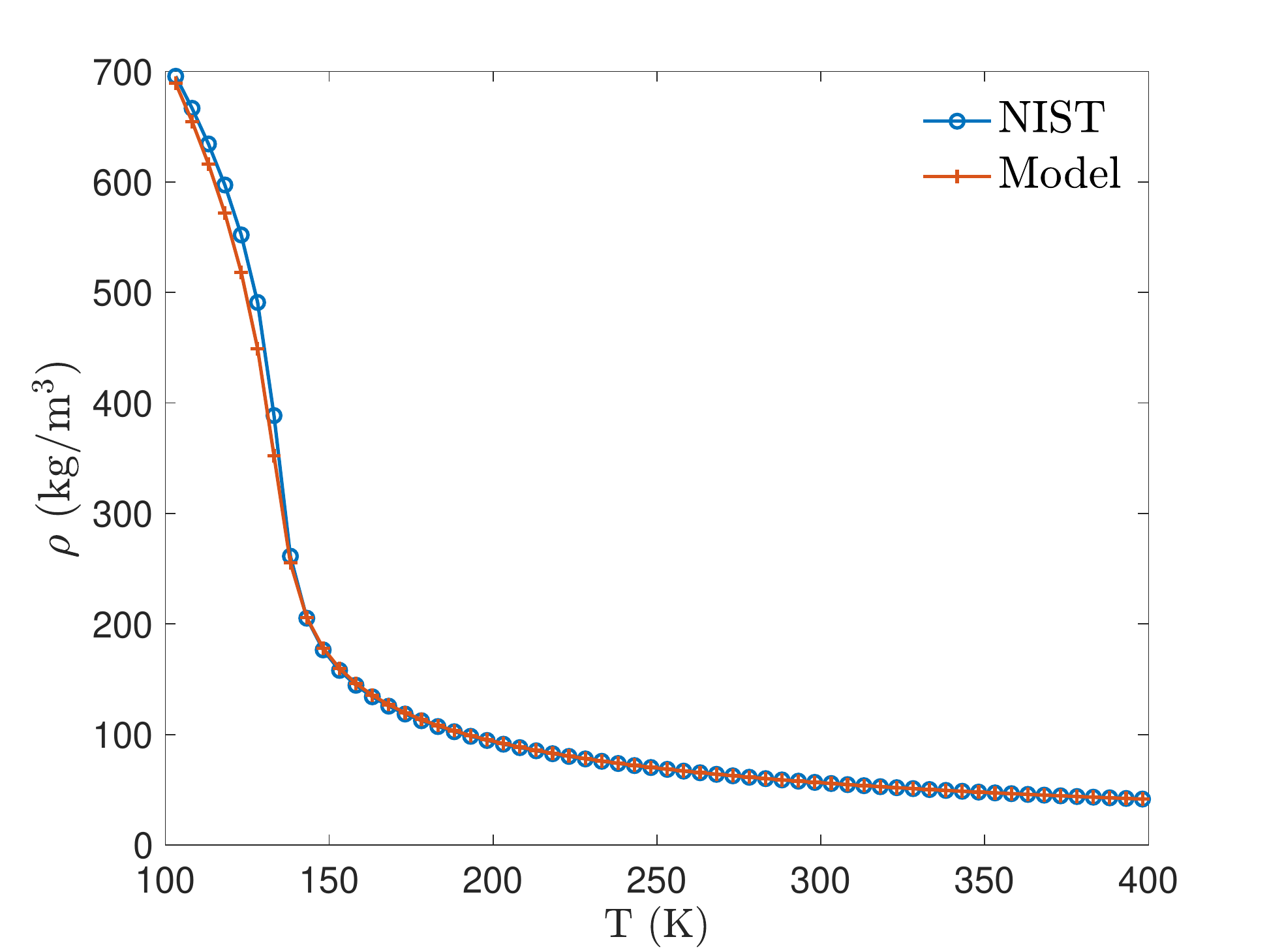}\includegraphics[width=9cm]{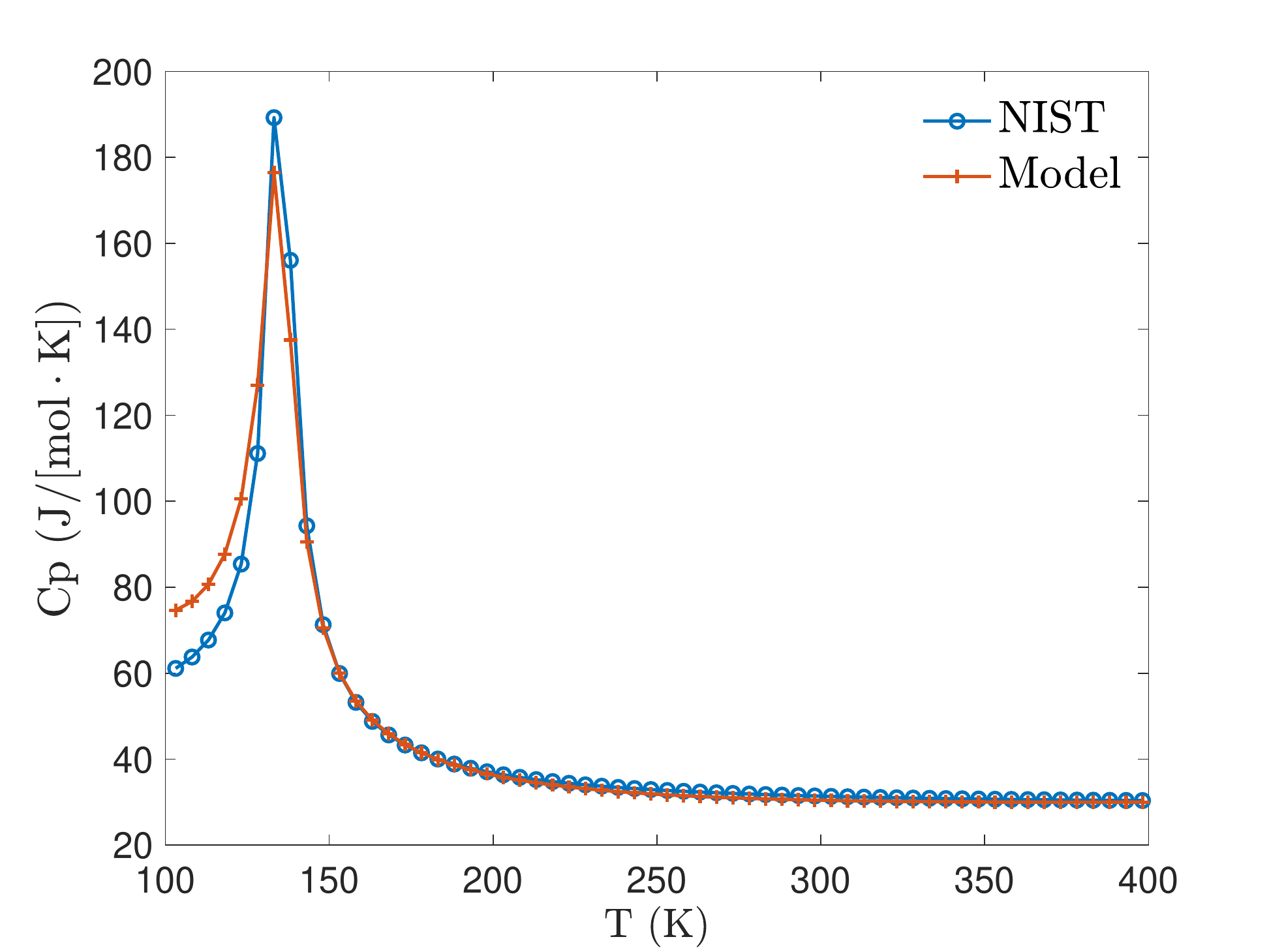}
\par\end{centering}

\begin{centering}
\qquad{}\qquad{}(a)\qquad{}\qquad{}\qquad{}\qquad{}\qquad{}\qquad{}\qquad{}\qquad{}\qquad{}\qquad{}(b)
\par\end{centering}

\begin{centering}
\includegraphics[width=9cm]{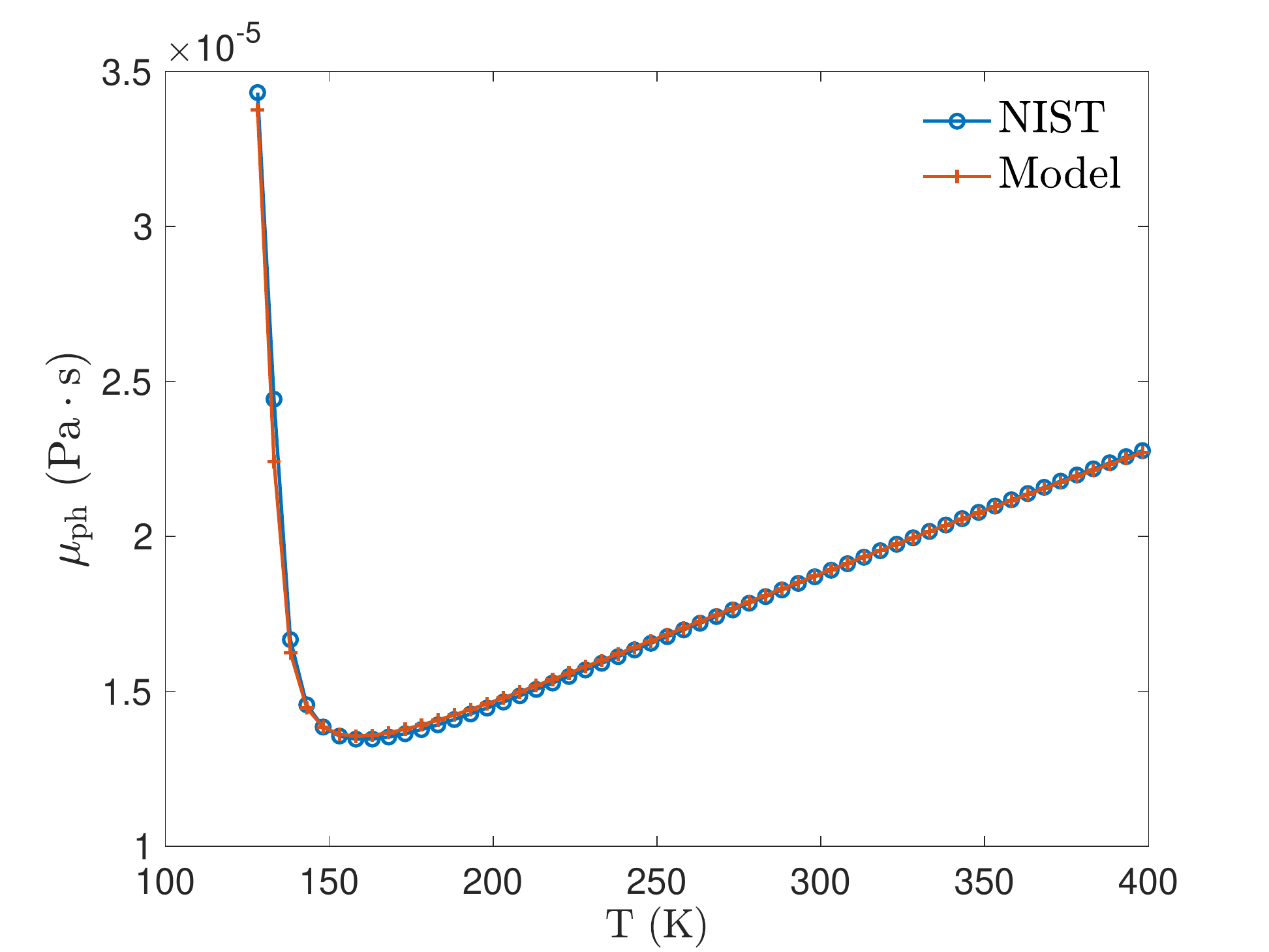}\includegraphics[width=9cm]{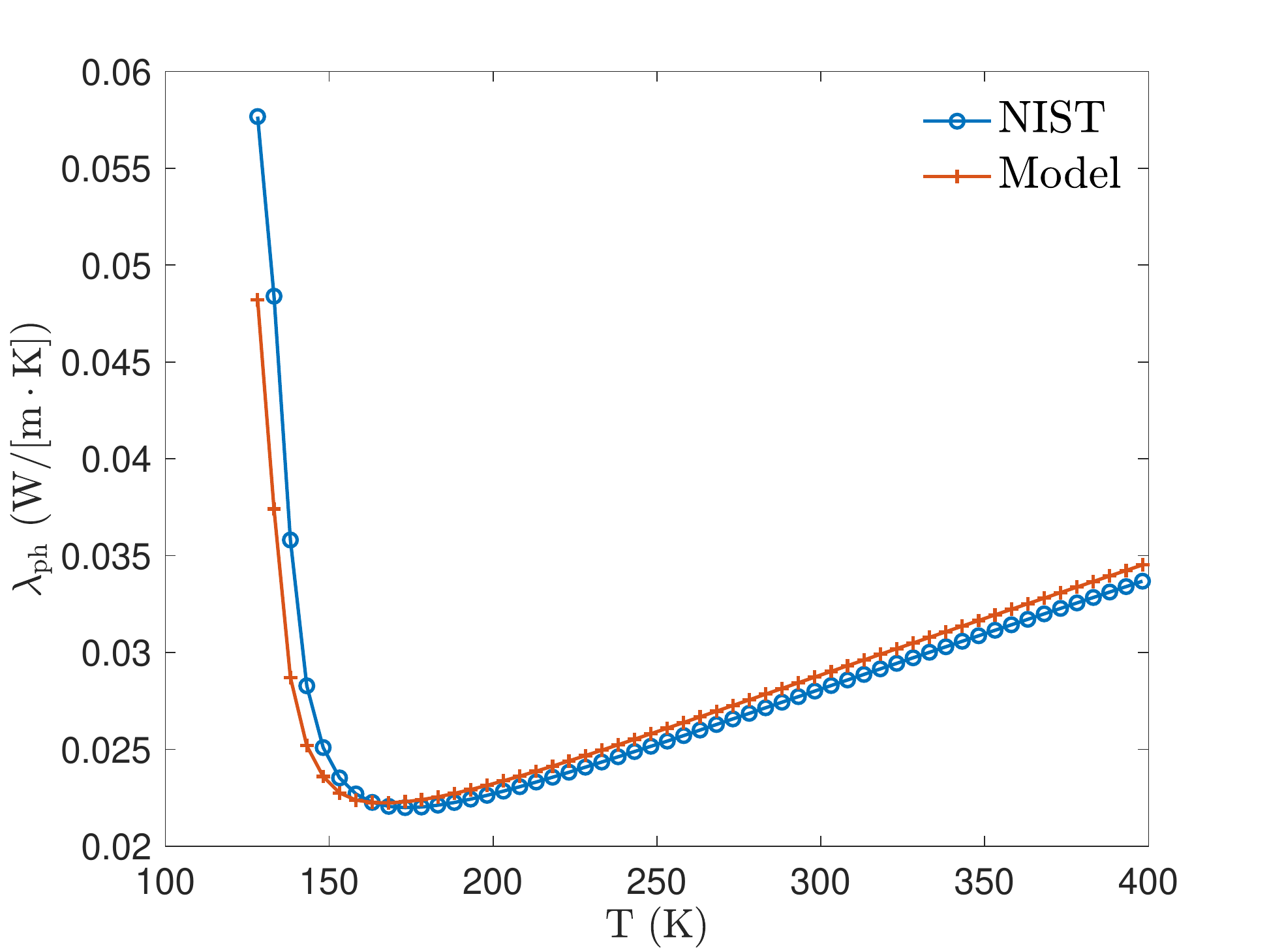}
\par\end{centering}

\begin{centering}
\qquad{}\qquad{}(c)\qquad{}\qquad{}\qquad{}\qquad{}\qquad{}\qquad{}\qquad{}\qquad{}\qquad{}\qquad{}(d)
\par\end{centering}

\caption{EOS and transport coefficients model comparison against NIST database
for pure Nitrogen at $50$ bar pressure. (a) Density, (b) Isobaric
heat capacity, (c) Viscosity, and (d) Thermal conductivity.\label{fig:NIST_vs_code_model}}
\end{figure}

\begin{figure}[!htb]
\begin{centering}
\includegraphics[width=9cm]{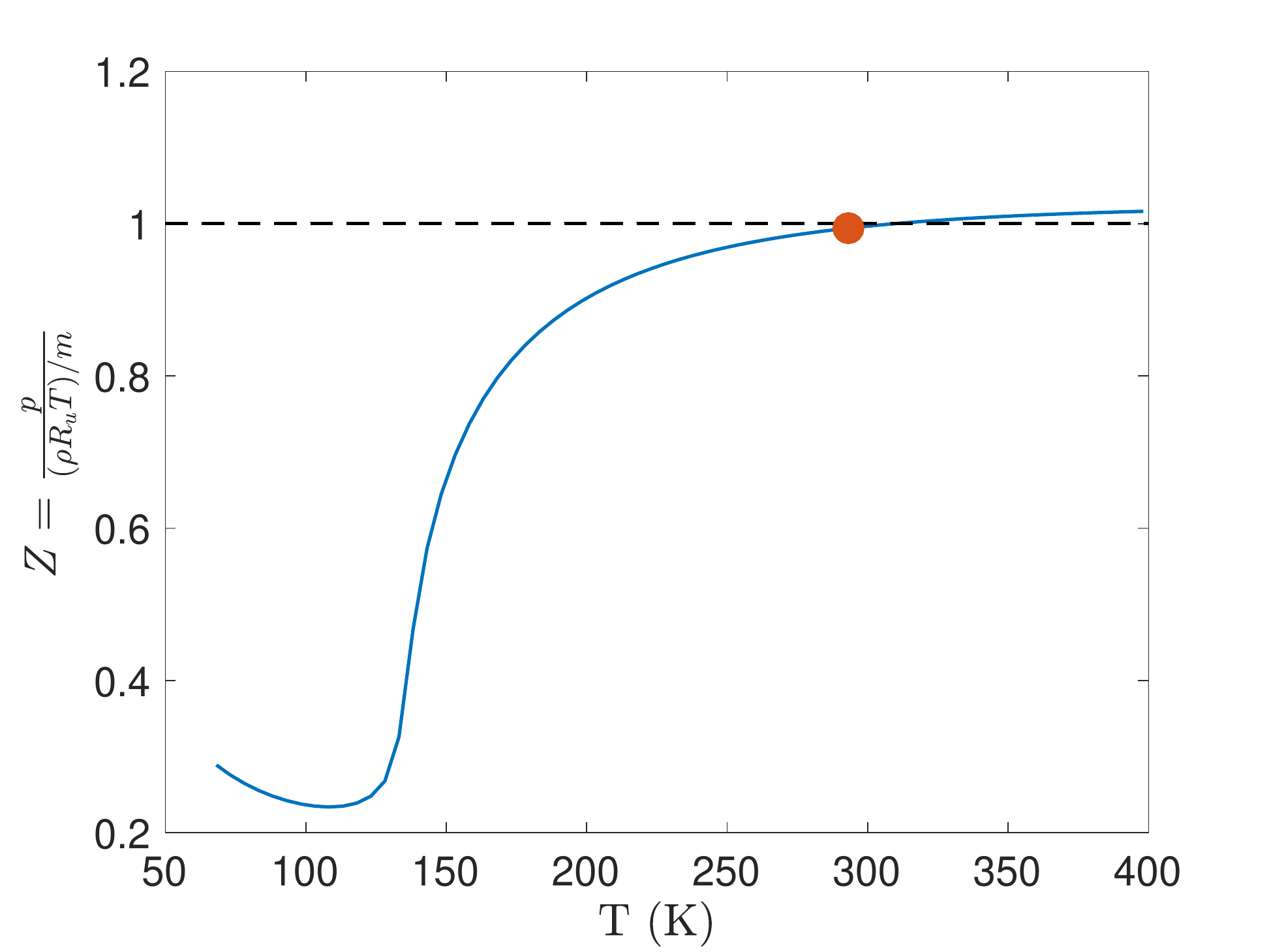}
\par\end{centering}

\caption{Compressibility factor of $\mathrm{N_{2}}$ at $50$ bar pressure.
Red marker denotes the chamber conditions for Case 2.\label{fig:comp_factor}}
\end{figure}

To examine the robustness of the above EOS and transport coefficient
models at supercritical conditions, Figure \ref{fig:NIST_vs_code_model}
compares the density, isobaric heat capacity, and the transport coefficients
$\mu_{\mathrm{ph}}$ and $\lambda_{\mathrm{ph}}$ obtained from the
models against the National Institute of Standards and Technology
(NIST) database \cite{lemmon2010nist} for $\mathrm{N_{2}}$ at a
pressure of 50 bar and temperatures ranging from 100 K to 400 K. The
supercritical temperature ($T_{c}$) of Nitrogen is 126.2 K. The transport
coefficient models are accurate only at supercritical temperatures
and, thus, the comparison only spans values of $T>T_{c}$. As evident,
the models have good agreement with the NIST database, showing their
validity at high-$p$ conditions encountered during Case 2 simulation.
Figure \ref{fig:comp_factor} shows the compressibility factor ($Z$),
indicating deviation from perfect-gas behavior, of pure Nitrogen for
a temperature range at $50$ bar pressure. The compressibility factor
at chamber conditions in Case 2 is $0.994$.

\noindent 
\begin{table}[!htb]
\begin{centering}
\begin{tabular}{ccccccc>{\centering}p{1cm}c}
\hline 
\multirow{2}{*}{Case} & \multirow{2}{*}{$N_{x}\times N_{y}\times N_{z}$} & $p_{\infty}$ & $T_{\mathrm{ch}}$ & $T_{\mathrm{inj}}$ & Number of & \multirow{2}{*}{$Re_{D}$} & \multirow{2}{1cm}{\centering{}$Ma_{0}$} & Inflow velocity\tabularnewline
 &  & (bar) & (K) & (K) & species &  &  & perturbation amplitude\tabularnewline
\hline 
\hline 
1 (atmP) & $320\times288\times288$ & 1 & 293 & 293 & 1 ($\mathrm{N_{2}}$) & 5000 & 0.6 & $0.004U_{0}$\tabularnewline
\hline 
2 (highP) & $320\times288\times288$ & 50 & 293 & 293 & 1 ($\mathrm{N_{2}}$) & 5000 & 0.6 & $0.004U_{0}$\tabularnewline
\hline 
\end{tabular}
\par\end{centering}

\caption{Summary of the conditions for the numerical simulations.\label{tab:Summary_of_cases}}
\end{table}

\noindent \vspace{-1.8cm}

\noindent 

\noindent 

\noindent \vspace{-1.0cm}

\section{Numerical details}

The spatial derivatives are approximated using the sixth-order compact
finite-difference scheme and time integration uses the explicit
fourth-order Runge-Kutta method. The outflow boundary in axial direction
and all lateral boundaries have sponge zones\cite{bodony2006analysis}
with subsonic non-reflecting outflow Navier-Stokes characteristic
boundary conditions (NSCBC)\cite{poinsot1992boundary} at the boundary
faces. Sponge zones at each outflow boundary have a width of $10\%$
of the domain length normal to the boundary face. The sponge strength
at each boundary decreases quadratically with distance normal to the
boundary. The performance of one-dimensional NSCBC\cite{poinsot1992boundary}
as well as its three-dimensional extension\cite{lodato2008three}
by inclusion of transverse terms were also evaluated without the sponge
zones; they permit occasional spurious reflections into the domain,
therefore, the use of sponge zones was deemed necessary. To avoid
unphysical accumulation of energy at the highest wavenumber, resulting
from the non-dissipative spatial discretization, the conservative
variables are filtered every five time steps using an eighth-order
filter.

\begin{figure}[!htb]
\begin{centering}
(a) \includegraphics[width=10.5cm]{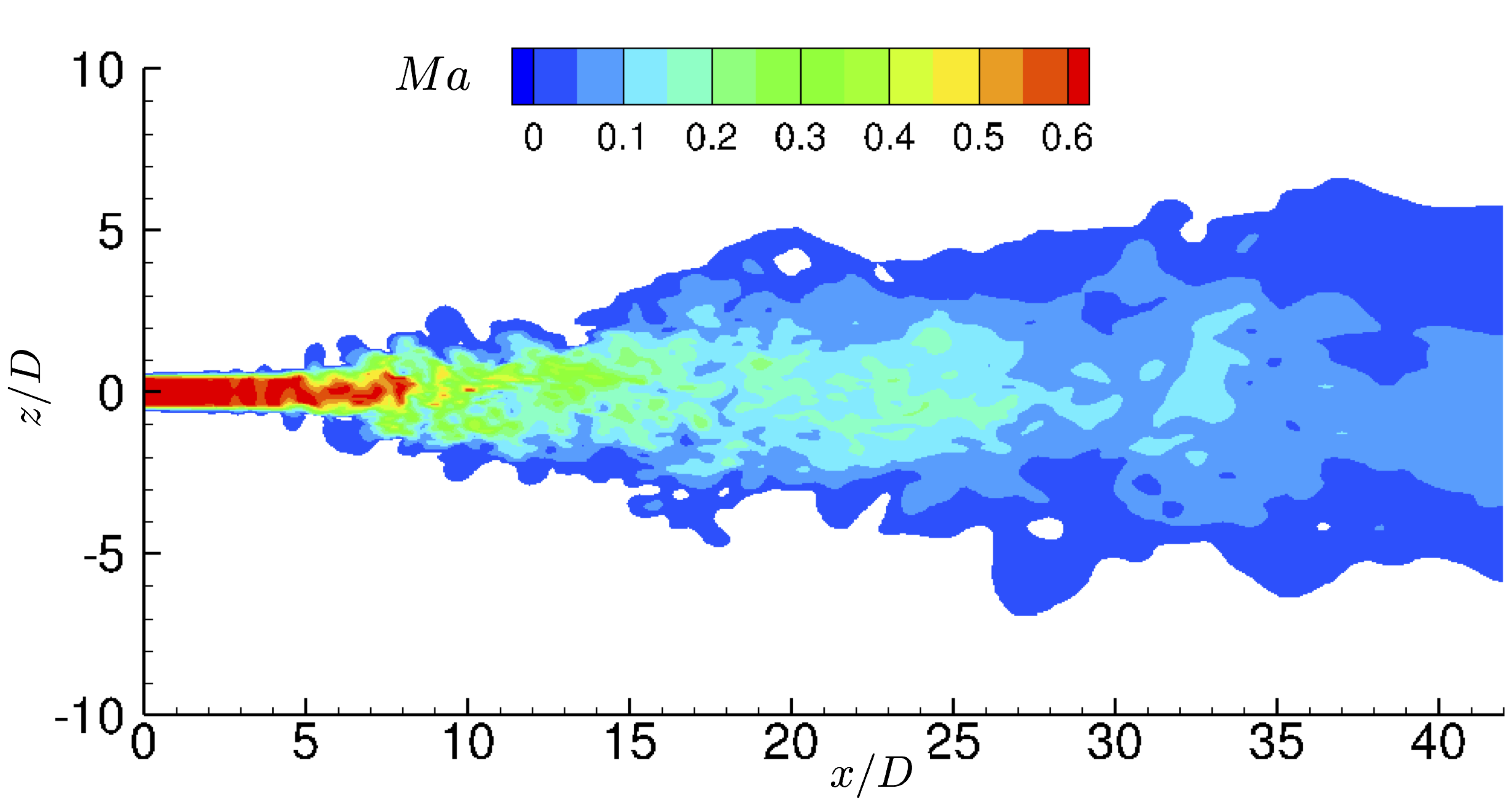}
\par\end{centering}

\begin{centering}
(b) \includegraphics[width=10.5cm]{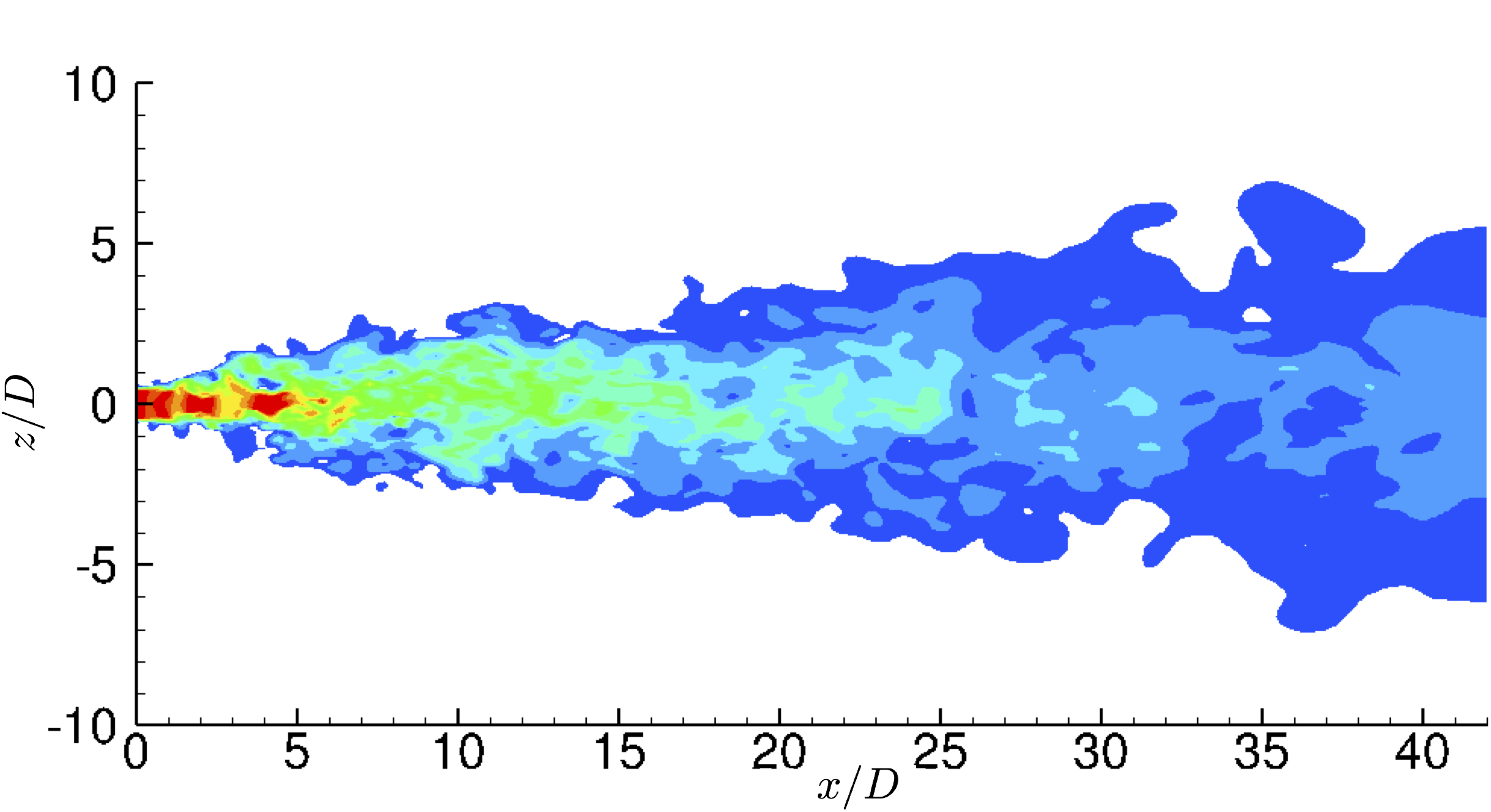}
\par\end{centering}

\caption{Instantaneous Mach number ($Ma$) field at $tU_{0}/D\approx2500$
in (a) Case 1 and (b) Case 2. Only values of $Ma\geq0.01$ are rendered.
Legend is the same for both plots.\label{fig:vel_scalar_contour}}
\end{figure}

The computational domain extends to $42D_{0}$ in the axial ($x$-)direction
and $20D_{0}$ in the $y$- and $z$-direction including the sponge
zones, as shown in Figure \ref{fig:vel_scalar_contour}. $320\times288\times288$
grid points are used in the $x\times y\times z$ direction, which
is twice the number of grid points in each direction used for DNS
by Boersma\cite{boersma2004numerical} at similar conditions as Case
1.

The axial grid resolution is chosen to resolve all spatial scales
overwhelmingly responsible for the dissipation. Following the approach
outlined in \cite{sharan2019numerical}, for a $Re_{D}=5000$ jet
simulation, the Kolmogorov length scale $\eta_{K}=0.0041\left(x-x_{0}\right)$,
where $x_{0}$ denotes the virtual origin. A stretched grid designed
accordingly is used for present simulations. Grid stretching is accounted
for by solving the governing equations in generalized coordinates
\cite{sharan2016time,sharan2018time}.

The velocity profile at the jet inflow plane is given by\cite{michalke1984survey}
\[
u(r)=\frac{U_{0}}{2}\left(1-\tanh\left[\frac{r-r_{0}}{2\theta_{0}}\right]\right),
\]
where the jet exit radius $r_{0}=D/2$ and the momentum thickness
$\theta_{0}=0.04r_{0}$ is assumed. The Mach number $Ma_{0}=U_{0}/c_{\infty}=0.6$
is specified, where $c_{\infty}$ denotes the speed of sound at ambient
conditions. Random perturbations with maximum amplitudes of $0.004U_{0}$
are superimposed on the inflow velocity profile to trigger jet flow
transition to turbulence. No perturbations are added to fields other
than velocity.

\vspace{-0.2cm}

\noindent 

\noindent 
\begin{figure}[!htb]
\begin{centering}
\includegraphics[width=9cm]{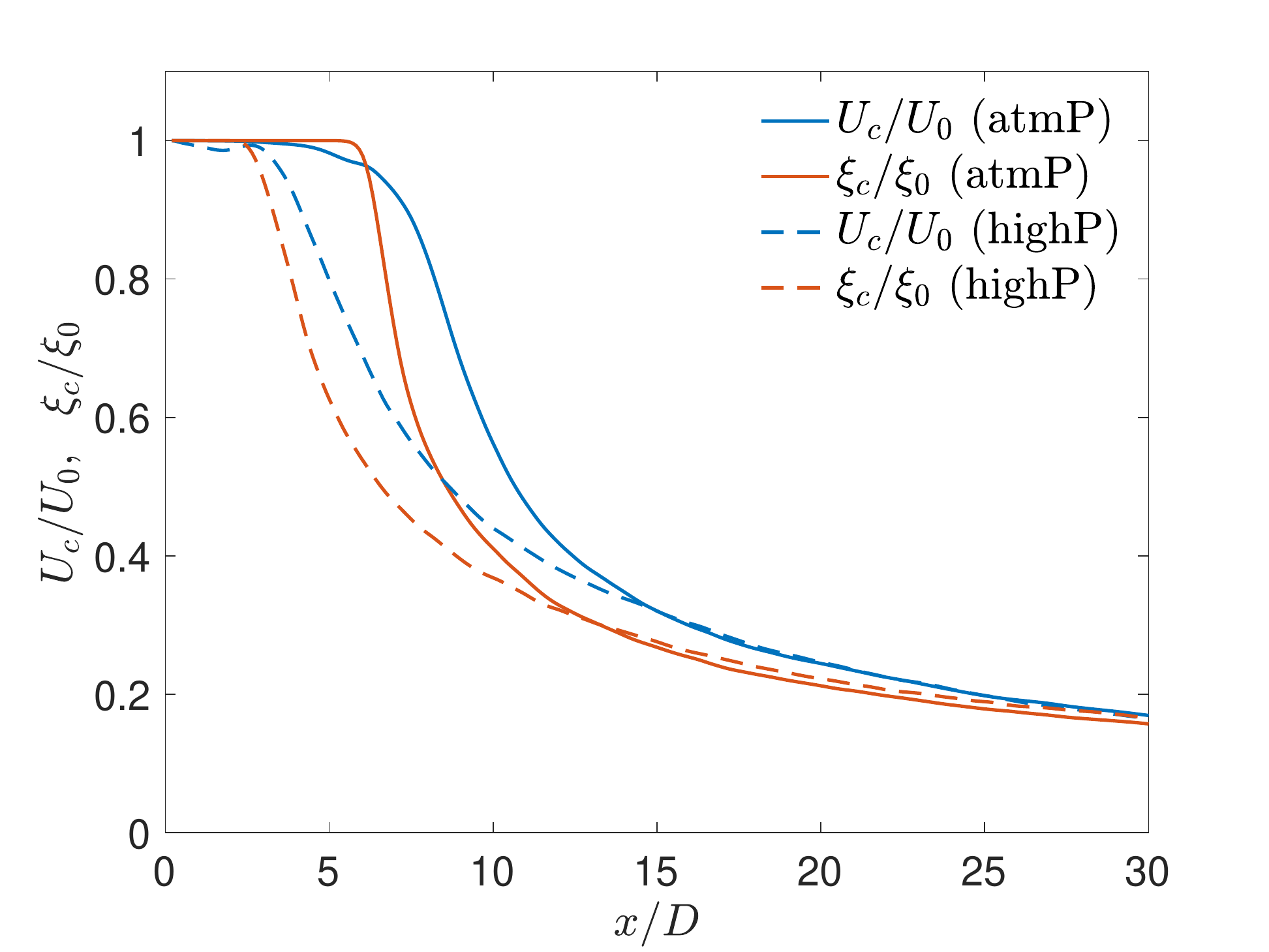}\includegraphics[width=9cm]{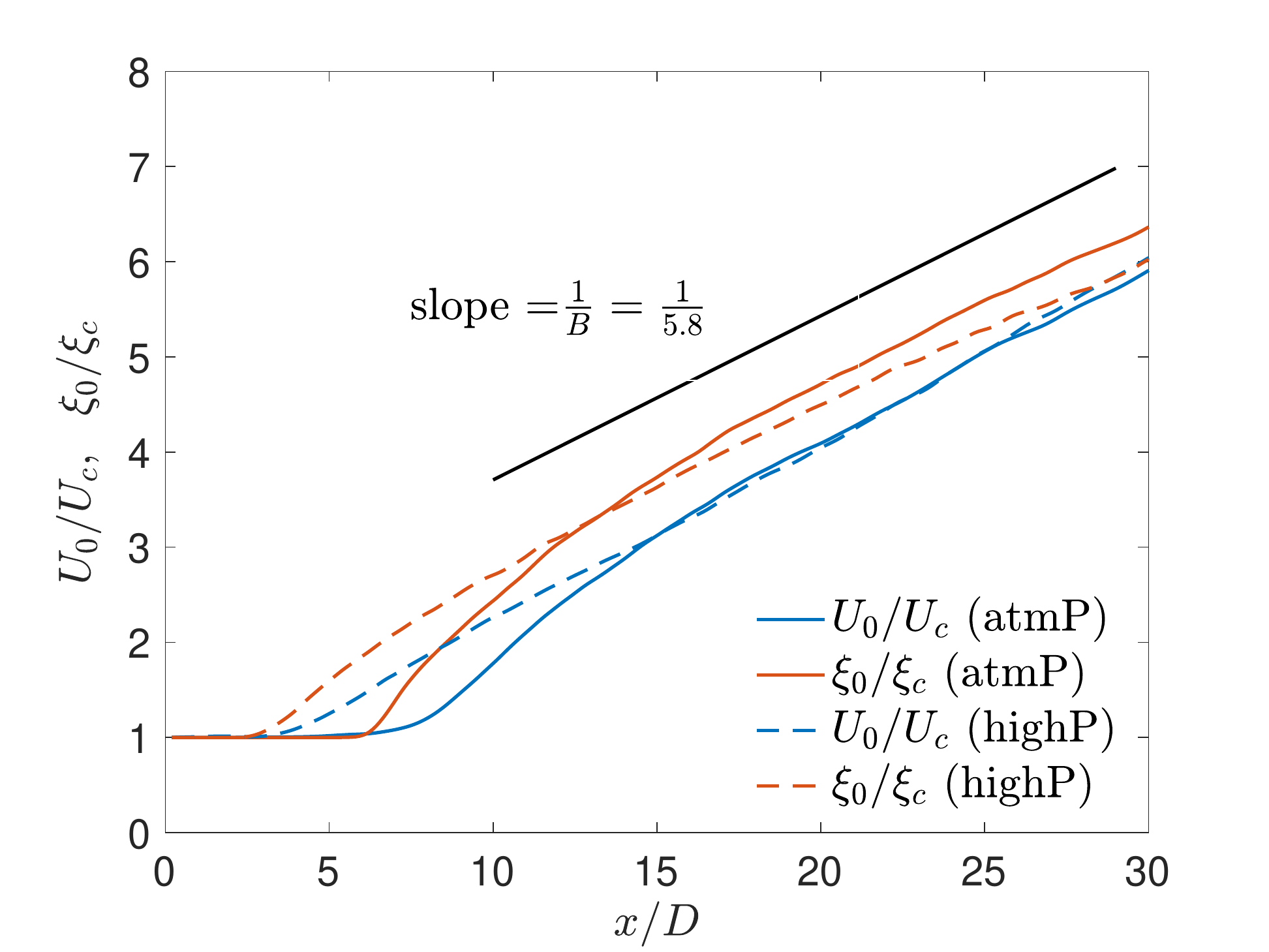}
\par\end{centering}

\begin{centering}
\qquad{}\qquad{}(a)\qquad{}\qquad{}\qquad{}\qquad{}\qquad{}\qquad{}\qquad{}\qquad{}\qquad{}\qquad{}(b)
\par\end{centering}

\caption{Case 1 and 2 comparison showing (a) the time-averaged centerline velocity
($U_{c}$) and scalar ($\xi_{c}$) values normalized with the jet
exit values $U_{0}$ and $\xi_{0}$ as a function of axial distance
and (b) the inverse of the time-averaged centerline values showing
linear decay asymptotically with axial distance.\label{fig:jet_centerline_vel_decay}}
\end{figure}

\noindent 
\begin{figure}[!htb]
\begin{centering}
\includegraphics[width=9cm]{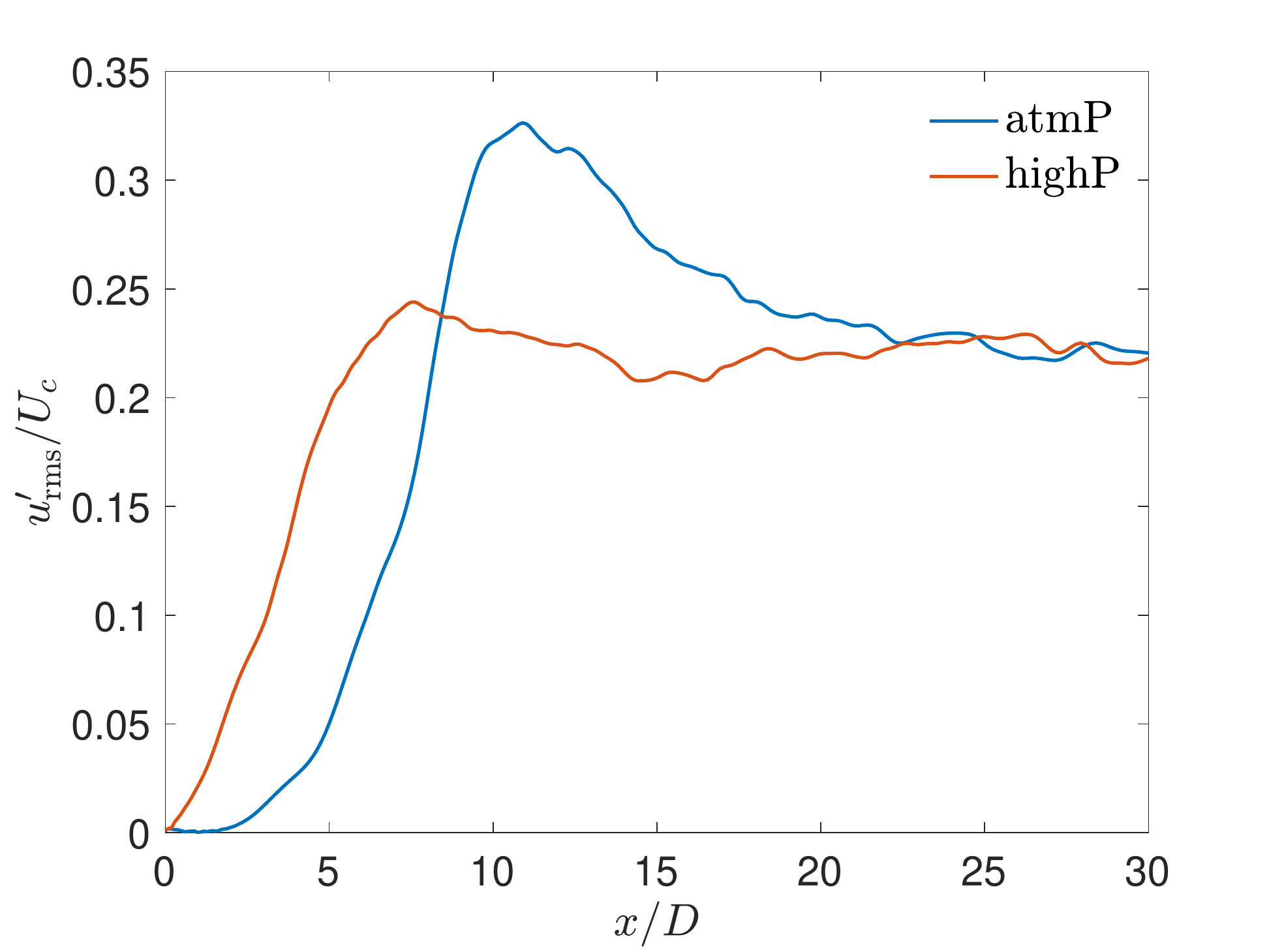}\includegraphics[width=9cm]{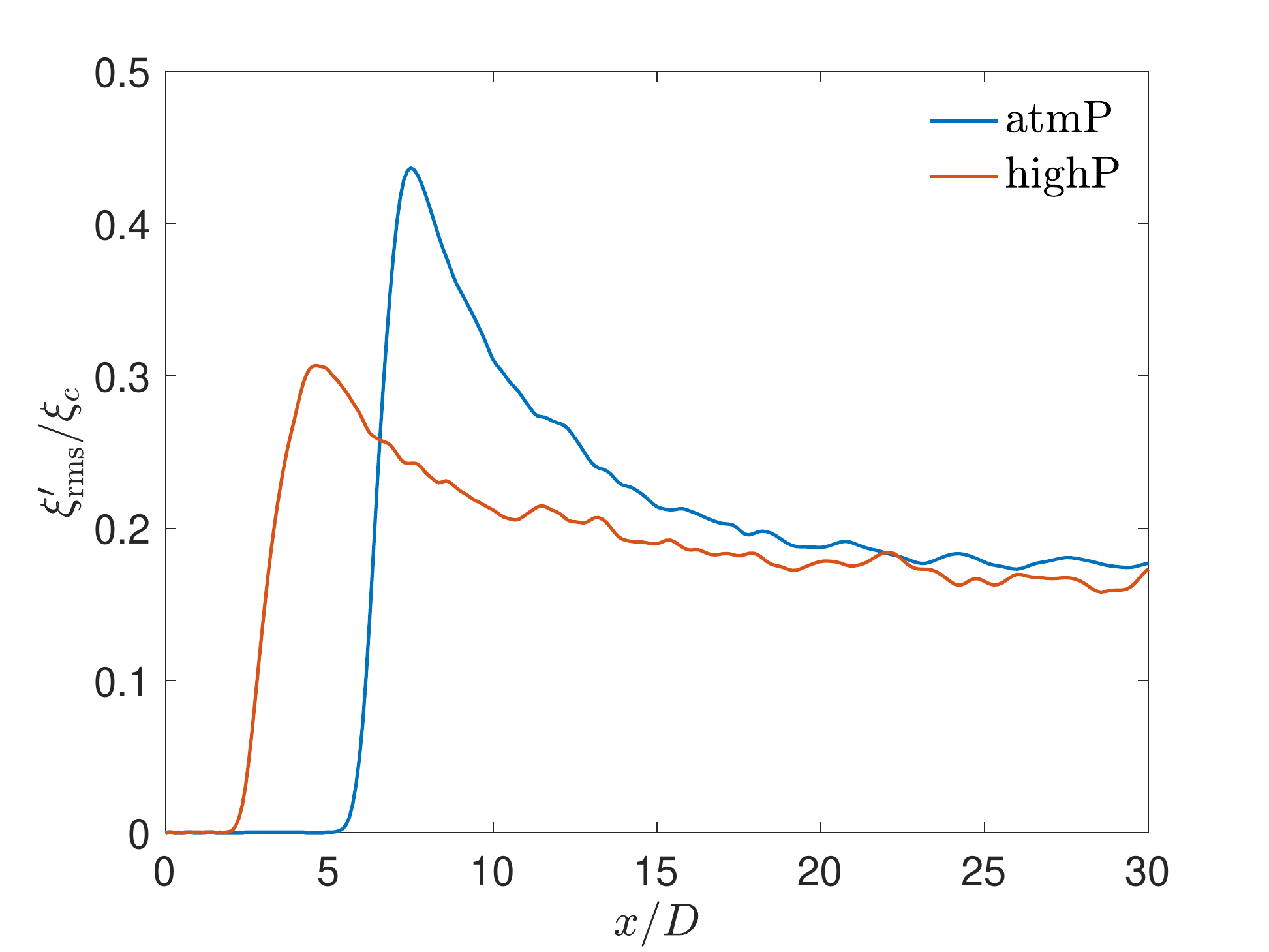}
\par\end{centering}

\begin{centering}
\qquad{}\qquad{}(a)\qquad{}\qquad{}\qquad{}\qquad{}\qquad{}\qquad{}\qquad{}\qquad{}\qquad{}\qquad{}(b)
\par\end{centering}

\caption{Centerline root-mean-square (a) axial velocity and (b) scalar fluctuation
comparison between Case 1 and 2.\label{fig:rms_fluc}}
\end{figure}

\noindent \vspace{-2.3cm}

\section{Jet flow results and discussion}

Table \ref{tab:Summary_of_cases} summarizes the conditions for the
numerical simulations considered in this study. In both cases, the
injected and chamber fluid temperatures and pressures are the same,
therefore, the jet injects into a chamber that is as dense as the
injected fluid. In other words, Case 1 and 2 represent jets at same
temperature and exit velocity, but in Case 1 a near-atmospheric pressure
jet is injected into similar chamber conditions and in Case 2 a $50$
bar pressure jet is injected into similar chamber conditions. Physically,
if the former jet has a $Re_{D}$ of 5000, the latter jet, based on
the density and viscosity at $50$ bar pressure and same exit velocity
$U_{0}$ and diameter $D$ as the velocity and length scale, will
have $Re_{D}\sim240,000$. A DNS at such $Re_{D}$ is infeasible,
therefore, in this study $Re_{D}$ is fixed to 5000 for both cases;
the $Re_{D}=5000$ condition is enforced using a computational viscosity
$\mu$ that is calculated by scaling the physical viscosity by a factor
$\mathcal{F}$, as discussed in Section \ref{sub:governing_eqn}.

$\mathcal{F}\approx6.5$ ($\mu_{R}=\rho_{0}U_{0}D/Re_{D}=1.136\times10^{-4}\thinspace\mathrm{Pa.s}$
and $\mu_{\mathrm{ph},0}=1.757\times10^{-5}\thinspace\mathrm{Pa.s}$)
for Case 1, near atmospheric pressure, and $\mathcal{F}\approx309.4$
($\mu_{R}=5.715\times10^{-3}\thinspace\mathrm{Pa.s}$ and $\mu_{\mathrm{ph},0}=1.847\times10^{-5}\thinspace\mathrm{Pa.s}$)
for Case 2 at 50 bar pressure. The factor $\mathcal{F}$ is higher
in Case 2 because of the higher density $\rho_{0}$ at $50$ bar that
requires a higher $\mu_{R}$ for a given Reynolds number $Re_{D}$.
The physical viscosity $\mu_{\mathrm{ph},0}$, on the other hand,
remains relatively unchanged with increase in pressure. Simulating
the two jets at same Reynolds number by using a computational viscosity,
as explained above, results in the high-$p$ jet becoming unphysically
more viscous than the atmospheric-$p$ jet. The following comparisons
between the two cases must account for this effect.

Figure \ref{fig:vel_scalar_contour} shows the Mach number contours
at $tU_{0}/D\approx2500$ for both cases. The potential core is comparatively
shorter in the high-$p$ case, likely due to higher computational
viscosity $\mu$, from a higher value of factor $\mathcal{F}$, and
resulting momentum diffusion.

Figure \ref{fig:jet_centerline_vel_decay} shows the decay of time-averaged
centerline velocity ($U_{c}$) and scalar concentration ($\xi_{c}$)
with axial distance. For a self-similar round jet with top-hat exit
velocity profile, the centerline velocity $U_{c}\left(x\right)$ is
given by the empirical relation\cite{hussein1994velocity}
\begin{equation}
\frac{U_{c}\left(x\right)}{U_{0}}=\frac{B}{\left(x-x_{0}\right)/D},\label{eq:centerline_vel}
\end{equation}
where $B$ is a constant and $x_{0}$ denotes the virtual origin.
As evident from Figure \ref{fig:jet_centerline_vel_decay}(b), downstream
of the potential core collapse, both the time-averaged centerline
velocity and scalar concentration decays as inverse of the axial distance,
where the rate of decay, given by $B$, is within experimentally observed
range of values \cite{hussein1994velocity}. The scalar concentration
begins to decay upstream of velocity, consistent with the observation
of Lubbers \textit{et al.} \cite[see Figure 6]{lubbers2001simulation}
for a passive scalar diffusing at unity Schmidt number. Despite the
difference in axial location where the velocity or scalar decay begins
between the near-atmospheric-$p$ and high-$p$ cases, the profiles
match asymptotically with axial distance. At the time of reporting,
the scalar statistics for the high-$p$ case did not fully converge;
we expect the dashed red line in Figure \ref{fig:jet_centerline_vel_decay}
to asymptotically converge to the solid red line (like the velocity
field shown in blue).

Figure \ref{fig:rms_fluc} shows the centerline root-mean-square (r.m.s.)
axial velocity and scalar concentration fluctuation, denoted by $u'_{\mathrm{rms}}$
and $\xi'_{\mathrm{rms}}$, respectively, normalized by the time-averaged
centerline values. The centerline r.m.s. values are calculated from
\[
u'_{\mathrm{rms}}=\left(\overline{\left(u-U_{c}\right)^{2}}\right)^{1/2}=\left(\overline{u^{2}}-U_{c}^{2}\right)^{1/2}
\]
and a similar expression for scalar concentration. The overbar, $\overline{\bullet}$,
denotes a time average at the centerline. The r.m.s. velocity fluctuation
profiles in Figure \ref{fig:rms_fluc}(a) compare favorably with the
profile of Crow \& Champagne \cite[see Figure 13]{crow1971orderly}
and, similarly, the scalar fluctuation profiles in Figure \ref{fig:rms_fluc}(b)
compare well with the distributions of r.m.s. scalar fluctuation in
jets from smooth contraction nozzle shown in Mi \textit{et al}. \cite[see Figure 4(a)]{mi2001influence}.
As observed in Figure \ref{fig:jet_centerline_vel_decay}, despite
the differences in r.m.s. fluctuations in near-jet regions between
the high-$p$ and near-atmospheric-$p$ cases, the profiles match
asymptotically with axial distance. The fluctuation magnitude signifies
turbulence intensity, which is negligible in the potential core, increases
sharply with collapse of the core, and asymptotes downstream to a
constant value as the flow becomes self-similar. The rise in fluctuation
profiles at shorter axial distance in case of high-pressure jet shows
a shorter potential core. Relatively lower peak fluctuation amplitudes
in high pressure results is likely a manifestation of the higher computational
diffusivity from higher value of $\mu$.

\section{Conclusions}

Single-species turbulent jet simulations at different chamber conditions
are performed as part of an effort to understand fuel injection and
fuel-oxidizer mixing at high pressures. Two cases with near-atmospheric
and supercritical chamber pressures, respectively, are analyzed while
keeping inflow, boundary conditions and initial condition identical.
The equation of state and transport coefficient models are chosen
specific to the two conditions. The transport property models at supercritical
conditions are validated against the NIST database values. To avoid
subgrid-scale model errors and its, often difficult to predict, interactions
with thermodynamic calculations in high-Reynolds-number simulations,
DNS at $Re_{D}$ of $5000$ is performed. Profiles of centerline velocity
and scalar concentration mean and r.m.s. fluctuations show favorable
agreement with the experimental data at atmospheric-$p$ conditions.
For comparisons in this study, the near-atmospheric-$p$ and high-$p$
jets were simulated at the same Reynolds number by using a computational
viscosity adjusted accordingly. Matching the Reynolds number in such
a manner results in the high-$p$ jet becoming more viscous than the
atmospheric-$p$ jet, thus, obscuring mixing statistics' one-to-one
comparisons. An alternative is to match the inflow momentum instead
of the Reynolds number $Re_{D}$, keeping the viscosity physical.
This alternative will be a subject of future investigation.

\section{Acknowledgements}
This work was supported by the Army Research Office under the direction of Dr. Ralph Anthenien. The computational resources were provided by the NASA Advanced Supercomputing at Ames Research Center under the $\mathrm{T^3}$ program directed by Dr. Michael Rogers.

\printbibliography

\end{document}